\documentclass[aps,prb,twocolumn,superscriptaddress,floatfix]{revtex4}

\usepackage{bm}
\usepackage{amsmath}
\usepackage{amssymb}
\usepackage{graphicx}
\usepackage{color}
\usepackage{soul,xcolor}
\setstcolor{red}
\usepackage{ulem}
\usepackage{cancel}

\newcommand\Ccancel[2][red]

\usepackage{xcolor}\definecolor{ao}{rgb}{0.0,0.0,1.0}

\definecolor{br}{rgb}{1.0, 0.22, 0.0}

\usepackage{epsfig}

\input{epsf}

\begin{document}
\title{Heat currents in electronic junctions driven by telegraph noise}

\author{O. Entin-Wohlman}
\email{oraentin@bgu.ac.il}
\affiliation{Raymond and Beverly Sackler School of Physics and Astronomy, Tel Aviv University, Tel Aviv 69978, Israel}
\affiliation{Physics Department, Ben Gurion University, Beer Sheva 84105, Israel}

\author{D. Chowdhury}
\email{debashreephys@gmail.com}
\affiliation{Physics Department, Ben Gurion University, Beer Sheva 84105, Israel}

\author{A. Aharony}
\affiliation{Raymond and Beverly Sackler School of Physics and Astronomy, Tel Aviv University, Tel Aviv 69978, Israel}
\affiliation{Physics Department, Ben Gurion University, Beer Sheva 84105, Israel}

\author{S. Dattagupta}
\affiliation{
Jawaharlal Nehru Centre for  Advanced Scientific Research, Bangalore       560064, India}
\affiliation{
Bose Institute, Kolkata 700054, India}
\date{\today}

\begin{abstract}

The  energy and charge fluxes carried by electrons in a two-terminal junction subjected to  a random telegraph noise,  produced by a single electronic defect,  are analyzed. The telegraph processes are imitated by the action of a stochastic electric field that acts on the electrons in the junction. Upon averaging over all random events of the telegraph process, it is found that this  electric field  supplies, on the average,  energy to the electronic reservoirs, which is distributed unequally between them: the stronger is the coupling of the reservoir with the junction, the more energy it  gains. Thus the noisy environment can lead to a temperature gradient across an un-biased junction.

\end{abstract}

\maketitle

\section{Introduction}

\label{Intro}

Nanoscopic electronic devices
driven by time-dependent fields  (beside  being subjected to stationary voltages and/or  temperature gradients), are currently attracting considerable attention, due to the possibility to control quantum-coherent
charge  and heat dynamics  in the time domain.
Experimental endeavors
aimed to manipulate quantum capacitors, \cite{Glattli2006} flying single electrons \cite{Tarucha2016} and other charge excitations, \cite{Glattli2013}
or to perform fast thermometry,  \cite{Pekola2015,Campisi2015} are conspicuous examples.
These systems are also the  topic of numerous theoretical studies, in which the time-dependent sources are
oscillating  electric fields, \cite{Arrachea2007,Esposito2010,Liu2012,Arrachea2014,Splettstoesser2016}  periodic temperature variations,  \cite{Brandner2015}
 and periodic time-dependent hybridizations of the mesoscopic system and its reservoirs. \cite{Esposito2015A,Dare2016} [See  Ref. \onlinecite{Arrachea2016_EN} 
 for  an extensive review.]

\begin{figure}[htp]
\includegraphics[width=6cm]{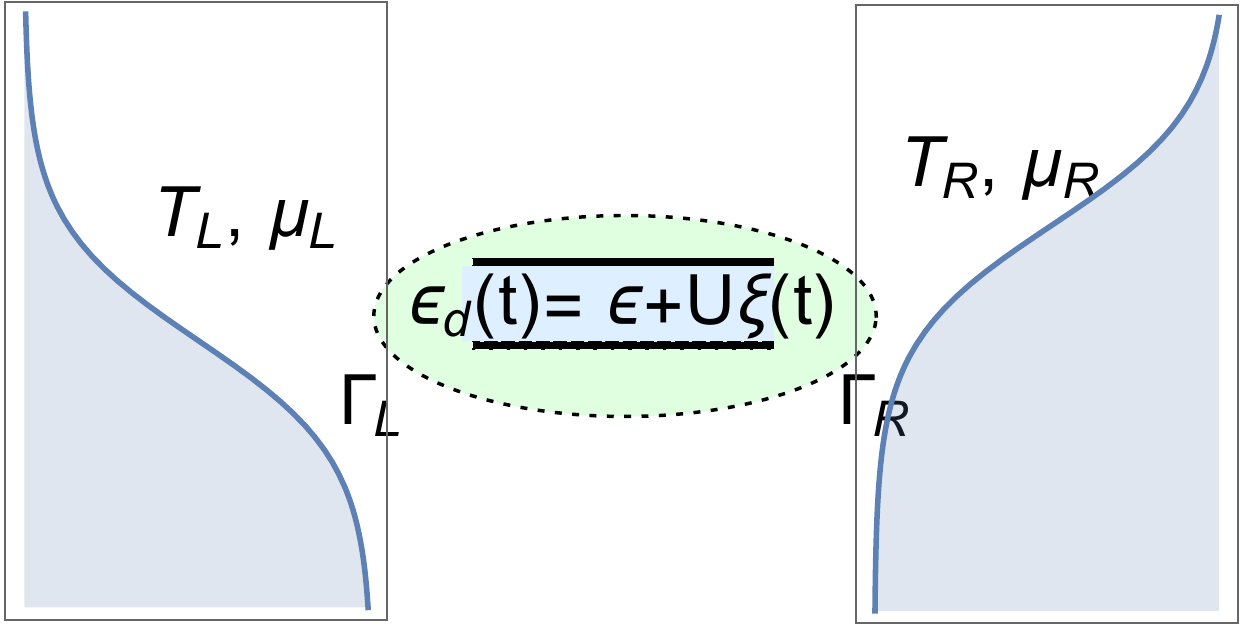}
\caption{Schematic picture of the model junction: a localized electronic level   is coupled to two electronic reservoirs, held at  chemical potentials $\mu_{L}$ and $\mu_{R}$, and at  temperatures $T_{L}$ and $T_{R}$, respectively. An electron residing on this level is subjected to a stochastic electric field. This is imitated by a stochastic time dependence of the level energy,  $\epsilon_{d}(t)$. The coupling with the reservoirs causes the level $\epsilon$ to become a resonance, of width $\Gamma=\Gamma_{L}+\Gamma_{R}$, where $\Gamma_{L},\Gamma_{R}$ are the partial widths. }
\label{Fig1}
\end{figure}

The precise definition of energy and heat  currents in the time domain in such junctions,  which comprise  nanoscale devices that accommodate a relatively small number of electrons coupled to
macroscopic reservoirs, is a fascinating theoretical subject. \cite{Pekola2014,Esposito2015}
A meticulous analysis can be carried out when the external force operating on the mesoscopic  system is slowly-varying in time,   \cite{Esposito2015,Nitzan2016}
particularly in the regime of  adiabatic quantum pumping,  \cite{Arrachea2007,Moskalets2004,Moskalets2009}
where proper response coefficients can be derived via the Kubo formula. \cite{Arrachea2016_93}

An intriguing feature concerning  heat transport in the time domain is the role played by the  energy flux associated  with the term in the Hamiltonian that couples the nanostructure with the bulky reservoirs. \cite{Pekola2014,Dare2016}
While the coupling part does not contribute directly to the particle flux, it does store energy momentarily, even when the hybridization between the nano system and the reservoirs does not vary with time.
The generic Hamiltonian in which the time-dependence is confined to the nano system (see Fig. \ref{Fig1}) is
\begin{align}
{\cal H}={\cal H}^{}_{\rm leads}+{\cal H}^{}_{\rm sys}(t)+{\cal H}^{}_{\rm tun}\ ,
\label{genH}
\end{align}
where ${\cal H}_{\rm tun}$ pertains to the coupling between the nano system [whose Hamiltonian is ${\cal H}_{\rm sys}(t)$] and the reservoirs (described by ${\cal H}_{\rm leads}$). The operator of  the particles' current, say of the left lead,   is simply the time-derivative of the particles' number operator in that lead. On the other hand, the operator of the total energy flux is the time derivative of the total Hamiltonian.  It comprises the energy flux associated with the particles in each of the leads, the one associated with the coupling, and the one arising from the time-derivative of ${\cal H}_{\rm sys}$. The latter consists of the energy flux of the particles residing on the nano system, and the explicit time derivative of the applied force, which amounts to the power supplied to the junction by the time-dependent field. Thus, each of these fluxes has an intuitive meaning except perhaps for the flux arising from the tunneling term.
While being a negligible contribution in macroscopic systems, it becomes comparable to the other energy fluxes in the nanoscale, where the surface to volume ratio is not vanishingly small.
The proposal to regard this energy flux as part of the time-dependent heat current of the corresponding reservoir,  \cite{Arrachea2014,Arrachea2016_94,Arrachea2017} based on a comparison  between the Green's function and the scattering approaches for calculating averaged quantities of noninteracting electrons,  \cite{Arrachea2006}     is still debated. \cite{Esposito2015A,Lopez2015,
Nitzan2016,Nitzan2016A}

The lion's share of the theoretical literature  on thermoelectric phenomena in mesoscopic junctions  in the time domain  focuses on periodic ac fields, preferably of frequencies  low compared to
the inverse   time it takes the  carriers to traverse the sample. \cite{Moskalets2004}
Here we consider  charge and energy currents in a nano device subjected to a time-dependent source which is   a noisy environment, described by
the so-called telegraph process, or telegraph noise. \cite{Blume1968,Dattagupta}
Telegraph noise is believed to result from the (almost unavoidable)  presence  of defects with internal degrees of freedom (coined `elementary fluctuators') that have two (or more) metastable configurations and can switch between them due to their interaction with a thermal bath (of their own).  In many cases, the fluctuations are due to the dynamics of charge carriers trapped at the defects.  \cite{Rogers1985,Hitachi2013,Kuhlmann2013,Lachance2014,BarJoseph2014}
This picture of a noisy environment  has been widely exploited at the time  to study
decoherence and dephasing of qubits. \cite{Tokura2003,Marquardt2008,Cheng2008,Bergli2009,Yurkevich2010,Aharony2010}

Charge fluctuations as the ones that give  rise to telegraph noise  result in fluctuating electric potentials. \cite{Sanchez2010} The  transport of spinless electrons through a junction subjected to  a stochastic potential (or field) due to telegraph noise  has been analyzed in various regimes of the noise characteristics,  \cite{Galperin1994,Gurvitz2016} but to our best knowledge, energy fluxes in such junctions were not discussed. An exception are 
 Ref. \onlinecite{Sanchez2017}
that discusses the thermal transport,
Ref. \onlinecite{Kubala2017} that focuses on the thermopower in a junction subjected to electromagnetic environment.

Here we present a detailed study of the electrons' energy fluxes in this configuration, carried out for
 the simplest (but realizable \cite{Linke2017})   junction, that of a single localized level (referred to below as ``quantum dot") attached to two reservoirs of spinless electrons. In addition to the possible relevance of this study to the influence of environments on electronic and thermoelectric  transport, we also aim at a  comparison of these effects with those of periodically oscillating fields.
 
The effect of the telegraph processes on the transport  is embedded in  the time-dependence of the energy on the   dot, $\epsilon_{d}(t)$, which  fluctuates randomly in time (see Fig. \ref{Fig1}). It turns out that, within the Keldysh formalism for time-dependent nonequilibrium Green's functions,  this simple geometry  is amenable to an analytical solution for the particle and the energy fluxes {\it in the time domain}. This happens
when it can be assumed that the densities of states in the electronic reservoirs that supply the electrons 
are faithfully represented by their value at the respective chemical potentials, an approximation coined ``the wide-band limit".
 \cite {Jauho1994,Odashima2017}
 We show that once  the time-dependent particle and energy fluxes are averaged over the processes of the telegraph noise they lose their dependence on time and become stationary. In particular, by examining the energy conservation conditions of the junction we obtain  the electric power supplied to the junction from the source of the  telegraph process. We find  that this power is not equally distributed between the two electronic reservoirs: rather the bigger part of the supplied power is absorbed by the reservoir which is connected more strongly to the dot. Thus, even when the electronic reservoirs  are  at identical temperatures and chemical potentials, the telegraph noise ``attempts" to create a temperature difference between them.
 It is interesting to invoke  in this context  Ref. \onlinecite{Arrachea2005}, which calculates the   charge current at zero bias and equal electronic temperatures  under the effect of a local ac field.   
 Averaged over a period, the resulting current is nonzero, as long as the  partial resonance widths ($\Gamma_{L}$ and $\Gamma_{R}$) on the dot are energy dependent (that is,  once  
the wide-band approximation is abandoned).

The  calculation of the electronic properties in the time domain,   by now rather standard, is well documented in the literature (mainly  with the aim of investigating the currents under the effect of oscillatory fields  [see, e.g., Ref. \onlinecite{Arrachea2016_EN}]). It is summarized briefly in Sec. \ref{Keldysh} and in  Appendix \ref{GF},  in a way which is particularly suitable for carrying out the average over the telegraph processes. The averaging procedure itself  is explained  in  Sec.  \ref{RTN} and in Appendix \ref{average}. 
After this somewhat technical detour, we return to the fluxes in our junction, and present the results for their stationary limits
in Sec. \ref{stationary}. Our conclusions are summarized in Sec. \ref{disc}.

\section{Energy and particle fluxes in the time domain}
\label{Keldysh}

As mentioned, our model system   consists of a localized level; hence,
\begin{align}
{\cal H}^{}_{\rm sys}=\epsilon^{}_{d}(t)d^{\dagger}_{}d\  ,
\label{Hsys}
\end{align}
where $d$ ($d^{\dagger}$) annihilates (creates) an electron on the level,  whose energy depends on time. This time dependence need not be specified in this section.   The dot is coupled to two electronic reservoirs  of spinless electrons  by tunneling amplitudes $V_{\bf k}$ and $V_{\bf p}$,
\begin{align}
{\cal H}^{}_{\rm tun}&=\sum_{\bf k}(V^{}_{\bf k}c^{\dagger}_{\bf k}d+{\rm Hc})+\sum_{\bf p}(V^{}_{\bf p}c^{\dagger}_{\bf p}d+{\rm Hc})\ ,
\label{Htun}
\end{align}
where $c^{}_{{\bf k}({\bf p})}$  ($c^{\dagger}_{{\bf k}({\bf p})}$)  are the annihilation (creation) operators for the electrons in the leads; the latter are modeled as free electron gases,
\begin{align}
{\cal H}^{}_{\rm leads}&=\sum_{\bf k}\epsilon^{}_{k}c^{\dagger}_{\bf k}c^{}_{\bf k}+
\sum_{\bf p}\epsilon^{}_{p}c^{\dagger}_{\bf p}c^{}_{\bf p}\ .
\end{align}
[We use the wave vector ${\bf k}$ (${\bf p}$) for the states on the left (right) lead.]

The various fluxes in this simple junction, in terms of the Keldysh Green's functions in the time domain, are as follows. 
The particle flux of the left lead, i.e., the rate of change of the number of particles there,  is
\begin{align}
I^{}_{L}(t)=\langle \frac{d}{dt}\sum_{\bf k}c^{\dagger}_{\bf k}c^{}_{\bf k}\rangle
=\sum_{\bf k}[V^{\ast}_{\bf k}G^{<}_{{\bf k}d}(t,t)-V^{}_{\bf k}G^{<}_{d{\bf k}}(t,t)]\ , 
\label{IL}
\end{align}
Here $G^{<}_{{\bf k}d}(t,t')=i\langle d^{\dagger}_{}(t')c^{}_{\bf k }(t)\rangle $ is the lesser Green's function; the angular brackets indicate the quantum average.  (This notation of the Green's function pertains to all others, e.g, $G_{dd}$ and $G_{{\bf k}{\bf p}}$.)  As usual, one expresses $I_{L}(t)$ in terms of the Green's functions on the dot, \cite{Jauho1994}
\begin{align}
I^{}_{L}(t)
=\int dt^{}_{1}[\Sigma^{}_{L}(t,t^{}_{1})G^{}_{dd}(t^{}_{1},t)-G^{}_{dd}(t,t^{}_{1})\Sigma^{}_{L}(t^{}_{1},t)]^{<}_{}\ ,
\label{ILd}
\end{align}
where $\Sigma^{}_{L}(t,t')$ is the self energy due to the tunnel coupling with the left lead,
\begin{align}
\Sigma^{}_{L}(t,t')=\sum_{\bf k}|V^{}_{\bf k}|^{2}g^{}_{\bf k}(t,t')\ ,
\label{SigL}
\end{align}
and $g_{\bf k}(t,t')$ is the Green's function of the decoupled left lead. The lesser Green's function of the product in Eq. (\ref{ILd}) is found by using the Langreth rules. \cite{Jauho1994} The particle flux of the right lead is derived from Eqs. (\ref{ILd}) and (\ref{SigL}) by interchanging $L\Leftrightarrow R$ and ${\bf k}\Leftrightarrow {\bf p}$. The flux of particles of the dot itself is then compensated by the sum of the two,
\begin{align}
I^{}_{d}(t)=\langle \frac{d}{dt}d^{\dagger}_{}d^{}_{}\rangle=-[I^{}_{L}(t)+I^{}_{R}(t)]\ ,
\end{align}
that is, particle number in the junction is conserved.

Explicitly, within the wide-band approximation,  the particle current Eq. (\ref{ILd}) in the time domain is
\begin{align}
I^{}_{L}(t)=2i\Gamma^{}_{L}\Big (\int\frac{d\omega}{2\pi} &f^{}_{L}(\omega)[G^{a}_{dd}(\omega,t)-G^{r}_{dd}(\omega,t)]\nonumber\\
&-G^{<}_{dd}(t,t)\Big )\ ,
\label{ILt}
\end{align}
where $
 \Gamma_{L(R)}=\pi|V^{}_{k_{\rm F}}|^{2}{\cal N}_{L(R)}$, $\Gamma=\Gamma^{}_{L}+\Gamma^{}_{R}$,
${\cal N}_{L(R)}$ is the   density of states of the left (right) lead at the respective Fermi energy,  and
\begin{align}
f^{}_{L(R)}(\omega)=[e^{(\omega-\mu^{}_{L(R)})/(k^{}_{\rm B}T^{}_{L(R)})}+1]^{-1}
\label{Fermi}
\end{align}
is the Fermi distribution there. The retarded and advanced Green's functions on the dot are
 \begin{align}
G^{r(a)}_{dd}(\omega,t)&=\mp i\int^{t}dt' e^{\pm i (t-t')(\omega\pm i\Gamma)
\mp i\int_{t'}^{t}d\tau\epsilon^{}_{d}(\tau)}
\ .
\label{GDRAW}
\end{align}
This result is derived in  Appendix \ref{GF}.     As shown there,
\begin{align}
G^{<}_{dd}(t,t)
&=\int\frac{d\omega}{2\pi}\Sigma^{<}_{}(\omega)\int^{t}dt'e^{-2\Gamma(t-t')}\nonumber\\
&\times[iG^{r}_{dd}(\omega,t')-iG^{a}_{dd}(\omega,t')]\ ,
\label{GLTT}
\end{align}
where
\begin{align}
\Sigma^{<}_{}(\omega)=2i[\Gamma^{}_{L}f^{}_{L}(\omega)+\Gamma^{}_{R}f^{}_{R}(\omega)]\ .
\label{siglew}
\end{align}
Equations (\ref{ILt}-\ref{siglew})  completely describe the particle current in the time domain.

We next turn to the  energy fluxes that flow in the time domain. First, there is the usual energy flux of the left lead,
\begin{align}
I^{E}_{L}(t)
=&\langle \frac{d}{dt}\sum_{\bf k}\epsilon^{}_{k}c^{\dagger}_{\bf k}c^{}_{\bf k}\rangle\nonumber\\
=&\sum_{\bf k}\Big (\epsilon^{}_{k}[V^{\ast}_{\bf k}G^{<}_{{\bf k}d}(t,t)-V^{}_{\bf k}G^{<}_{d{\bf k}}(t,t)]\Big )\ ,
\label{ILE}
\end{align}
and the analogous energy flux associated with the right reservoir, derived from Eq. (\ref{ILE})
by interchanging $L\Leftrightarrow R$ and ${\bf k}\Leftrightarrow {\bf p}$.
In terms of the Green's functions on the dot, this energy flux reads
\begin{align}
I^{E}_{L}(t)
=\int dt^{}_{1}[\Sigma^{E}_{L}(t,t^{}_{1})G^{}_{dd}(t^{}_{1},t)-G^{}_{dd}(t,t^{}_{1}\Sigma^{E}_{L}(t^{}_{1},t)]^{<}_{}\ ,
\label{ILEd}
\end{align}
where $\Sigma^{E}_{L}(t,t')$ is 
\begin{align}
\Sigma^{E}_{L}(t,t')=\sum_{\bf k}\epsilon^{}_{k}|V^{}_{\bf k}|^{2}g^{}_{\bf k}(t,t')\ .
\label{SigEL}
\end{align}
Explicitly, the rate of change of the energy in the left lead [see Eq. (\ref{ILE})], is
\begin{align}
I^{E}_{L}(t)&=2i\Gamma^{}_{L}\Big (
\int\frac{d\omega}{2\pi}\omega f^{}_{L}(\omega)[G^{a}_{dd}(\omega,t)-G^{r}_{dd}(\omega,t)]\nonumber\\
&-
\epsilon^{}_{d}(t)
G^{<}_{dd}(t,t)\Big )
\nonumber\\
&-i\Gamma^{}_{L}\int\frac{d\omega}{2\pi}\Sigma^{<}_{}(\omega)[G^{a}_{dd}(\omega,t)+G^{r}_{dd}(\omega,t)]\ .
\label{ILEWt}
\end{align}
Details of the derivation of this expression, in particular
the last  term on the right hand-side,  are given in Appendix \ref{GF}.
The particle and energy fluxes associated with the leads, i.e., Eqs. (\ref{ILd}) and (\ref{ILEd}), are all that is needed to investigate thermoelectric effects in stationary two-terminal electronic junctions. \cite{Benenti2017}

In the time domain, however,  there are two additional energy fluxes. The first,  which results from the temporal variation of the (left and right) tunneling Hamiltonians, Eq. (\ref{Htun}),
reads
\begin{align}
I^{E}_{{\rm tun}, L}&(t)
= \langle \frac{d}{dt}\sum_{\bf k}(V^{}_{\bf k}c^{\dagger}_{\bf k}d+{\rm Hc})\rangle
=\epsilon^{}_{d}(t)I^{}_{L}(t)-I^{E}_{L}(t)\nonumber\\
&+\sum_{{\bf k},{\bf p}}[V^{\ast}_{\bf k}V^{}_{\bf p}G^{<}_{{\bf k}{\bf p}}(t,t)-V^{}_{\bf k}V^{\ast}_{\bf p}G^{<}_{{\bf p}{\bf k}}(t,t)]\ ,
\label{ITLE}
\end{align}
(with an analogous expression for $I^{E}_{{\rm tun},R}$). The calculation of the last term on the right hand-side is carried out in Appendix  \ref{GF}. From Eq. (\ref{kp}) one finds
\begin{align}
\label{ITLEt1}
&I^{E}_{{\rm tun},L}(t)=\epsilon^{}_{d}(t)I^{}_{L}(t)-I^{E}_{L}(t)
\\
&+2\Gamma^{}_{L}\Gamma^{}_{R}\int\frac{d\omega}{2\pi}[f^{}_{R}(\omega)-f^{}_{L}(\omega)]
[G^{a}_{dd}(\omega,t)+G^{r}_{dd}(\omega,t)]
\ .
\nonumber
\end{align}
The second ``new" energy flux is the one that comes from the temporal variation of the dot's Hamiltonian,  Eq. (\ref{Hsys}),
\begin{align}
I^{E}_{d}(t)&=
\langle \frac{d}{dt}\epsilon^{}_{d}(t)d^{\dagger}_{}d\rangle
=-i\frac{d\epsilon^{}_{d}(t)}{dt}G^{<}_{dd}(t,t)+\epsilon^{}_{d}(t)I^{}_{d}(t)\ .
\label{IDE}
\end{align}
The first term on the right hand-side of Eq. (\ref{IDE}) results from the explicit time-dependence of the localized energy, i.e., it is due to time-dependent electric potential acting on the dot. Since the electronic occupation on the dot, \cite{Jauho1994} $Q_{d}(t)$, is
\begin{align}
Q^{}_{d}(t)=-iG^{<}_{dd}(t,t)\ ,
\label{Q}
\end{align}
this term expresses the power supplied to the system by the field. We  denote this power by $P_{d}(t)$, with
\begin{align}
P^{}_{d}(t)=Q^{}_{d}(t)\frac{d\epsilon^{}_{d}(t)}{dt}\ .
\label{P}
\end{align}
Combining Eqs. (\ref{ILE}), (\ref{ITLE}), and (\ref{IDE}), one finds
\begin{align}
I^{E}_{L}(t)+I^{E}_{R}(t)+I^{E}_{{\rm tun},L}(t)+I^{E}_{{\rm tun},R}(t)+I^{E}_{d}(t)=P^{}_{d}(t)\ ,
\end{align}
which expresses the energy conservation in the  junction.

\section{Average over the random telegraph process}
\label{RTN}

The localized energy on the dot in our junction fluctuates with time,
\begin{align}
\epsilon^{}_{d}(t)&=\epsilon +U\xi(t)\ ,
\label{ed}
\end{align}
where $\xi(t)$ describes  a random telegraph process.
Telegraph noise is an example of a dichotomous, stationary, and discrete process. Beginning at an initial time $t^{}_0$, the function $\xi(t)$ jumps instantaneously between the values $+1$ and $-1$ at random instants $t^{}_0< t^{}_{1}<t^{}_{2}<t^{}_{3}<\ldots<t$.
Each history of the system involves a certain sequence of times, at which the jumps occur.
In our calculation, we average the time-dependent fluxes (at time $t$) over all histories; this procedure amounts to averaging over all the time sequences, and over the two possible initial values of the random variable $\xi(t^{}_0)$. The average contains the case with no jump, the case with one jump at any intermediate time between the initial time $t_{0}$ and $t$,  the case with two jumps at any intermediate times $t^{}_0<t^{}_1<t^{}_2<t$, {\it etc}. The  average value of any physical quantity  depends only on the time difference, $t-t^{}_0$;  due to this property, the averages of the time-dependent  fluxes derived in Sec. \ref{Keldysh} become stationary.

In the simplest example, the telegraph noise is characterized by the {\it a priori} probabilities of the occurrence of $\xi=+1$ (or $-1$).
In the example of the elementary charge fluctuators, each fluctuator is assumed to be in one of two states; these states, which occur with probabilities $p^{}_+$ or $p^{}_-$, generate an electrostatic potential $+U$ or $-U$ on the electron which occupies the quantum dot. If the two states of the fluctuator have energies $0$ or $E_0>0$, and when their occupation is determined by their interaction with a separate heat bath of temperature $T$, then the probabilities are
given by the (normalized) Boltzmann factors,\cite{Galperin1994}
\begin{align}
p^{}_{\pm}(T)&=\frac{\exp[\pm E_0/(2k^{}_{\rm B}T)]}{2{\rm cosh}[E_0/(2k^{}_{\rm B}T)]}\ . 
\label{p}
\end{align}
The ``telegraph noise temperature" $T$ is an {\it effective} temperature, that models the probabilities $p^{}_\pm$. Since the model Hamiltonian does not include the interaction between the fluctuator and the dot, nor the back action from the electrons to the fluctuator, the system is not in equilibrium and there is no meaning to a comparison of $T$ with the temperatures of the electronic reservoirs. At zero ``telegraph-noise temperature", $T=0$, this example yields  $p^{}_+=1$, hence no fluctuations. The effect of the fluctuations then increases as $T$ is raised. Other models of the telegraph noise give similar results. With these probabilities, the average of $\xi(t)$ is independent of the time, 
\begin{align}
\overline{\xi}&=p^{}_{+}(T)-p^{}_{-}(T)={\rm tanh}[E_0/(2k^{}_{\rm B}T)]\ .
\label{overx}
 \end{align}
(We denote  an average over the telegraph  process   by an over bar, to distinguish it from the quantum average, which is indicated by angular brackets.) As $T\rightarrow\infty$, the average $\overline{\xi}$ tends to zero.
 
In addition to the probabilities $p^{}_\pm$, one also needs to specify the mean rate $\gamma$ at which the instantaneous jumps occur. Assuming detailed balance,  the probability per unit time to jump from $\xi=a(=\pm 1)$ to $\xi=b(=\mp 1)$ is $W^{}_{ab}=\gamma p^{}_b$, and the probability per unit time to stay in the state $a$ is $W^{}_{aa}=-\sum_{b\ne a}W_{ab}=-\gamma p^{}_{-a}$. The total rate for any jump is $\gamma=W^{}_{1,-1}+W^{}_{-1,1}$. It is expedient to present the four possible values of $W_{ab}$ in a matrix form,
\begin{align}
{\bf W}&=\gamma\left[\begin{array}{cc}-p^{}_- &\ \  p^{}_- \\ \ \ p^{}_+ & -p^{}_+ \end{array}\right ]\  .
\label{WW}
\end{align}

To average over the telegraph noise histories, it is convenient to define conditional averages:  the average of the function $F(t,t')$ under the assumption that $\xi(t')=a$ and $\xi(t)=b$ (with $t>t'$) is expected to depend only on $t-t'$, and is hence denoted by $F(t-t')^{}_{ab}$. The average over all histories and all initial and final values of $\xi$ is 
 \begin{align}\
 \label{av}
 \overline{F(t,t')}=\sum_{a,b}p^{}_aF(t-t')^{}_{ab}\ .
 \end{align}
 In particular, the conditional average probability that $\xi(t)=b$, given that $\xi(t')=a$, is the $2\times 2$ matrix ${\bf P}(t,t')$,  which solves the differential equation
 \begin{align}
 \frac{d}{dt}{\bf P}(t,t')={\bf W P}(t,t')\ ,
 \end{align}
 with the initial condition ${\bf P}(t,t')={\bf I}$, the $2\times 2$ unit matrix. The solution depends only on $t-t'$,  
 \begin{align}
 \label{PP}
 {\bf P}(t,t')&\equiv {\bf P}(t-t')=e^{{\bf W}(t-t')}\nonumber\\
 &={\bf I}+{\bf W}/\gamma-{\bf W}e^{-\gamma (t-t')}/\gamma\ .
 \end{align}

 We now demonstrate the averaging over the telegraph noise by considering  the particle flux on the left lead, $I^{}_L(t)$.  As seen from  Eq.~(\ref{ILt}), this average requires the averages over $G^{a}_{dd}(\omega,t)=[G^{r}_{dd}(\omega,t)]^*$, which also   determine the average over  $G^{<}_{dd}(t,t)$,  Eq. (\ref{GLTT}). To this end, we rewrite  Eq. (\ref{GDRAW}) as
 \begin{align}
 \label{Gadd}
G^{a}_{dd}(\omega,t)&= i\int^{t}dt' e^{- i (t-t')(\omega-\epsilon-U\overline{\xi}- i\Gamma)}X(t,t')\ ,
\end{align}
with the random part
\begin{align}
X(t,t')= e^{iU\int_{t'}^{t}d\tau[\xi(\tau)-\overline{\xi}]}
\ .
\label{x}
\end{align}
Since $dX(t,t')/dt=iU[\xi(t)-\overline{\xi}]X(t,t')$, integration yields \cite{Blume1968}
\begin{align}
X(t,t')= 1+iU\int_{t'}^{t}d\tau[\xi(\tau)-\overline{\xi}]X(\tau,t')\ .
\end{align}
The conditional average of this equation is expected to depend only on $t-t'$, and it obeys the equation
\begin{align}
&X(t-t')^{}_{ab}=P(t-t')^{}_{ab}\nonumber\\
&+iU\sum_{c}\int_{t'}^{t}d\tau X(\tau-t')^{}_{ac}[c-\overline{\xi}]P(t-\tau)^{}_{cb}\ .
\end{align}
In matrix form,
\begin{align}
{\bf X}(t-t')&={\bf P}(t-t')\nonumber\\
&+iU\int_{t'}^{t}d\tau {\bf X}(\tau-t')[\sigma^{}_z-\overline{\xi}{\bf I}]{\bf P}(t-\tau)\ ,
\label{Xttp}
\end{align}
where $\sigma^{}_z$ is the Pauli matrix. 
Equation (\ref{Xttp}) is conveniently solved by Laplace transforming it,    \cite{Blume1968}
\begin{align}
\tilde{\bf X}(s)=\tilde{\bf P}(s)+iU\tilde{X}(s)(\sigma^{}_z-\overline{\xi}{\bf I})\tilde{\bf P}(s)\ ,
\end{align}
where 
$\tilde{f}(s)=\int_0^\infty d\tau e^{-s\tau}f(\tau)$.  In particular, 
$\tilde{\bf P}(s)=[s{\bf I}-{\bf W}]^{-1}$, which results from Eq. (\ref{PP}).
The solution for the matrix $\widetilde{\bf X}(s)$ is 
\begin{align}
&\tilde{\bf X}(s)=[\tilde{\bf P}(s)^{-1}-iU(\sigma^{}_z-\overline{\xi}{\bf I})]^{-1}\nonumber\\
&=\frac{1}{D}\left[\begin{array}{cc}s+iU(\overline{\xi}+1)+\gamma p^{}_+ & \gamma p^{}_- \\ \gamma p^{}_+ & s+iU(\overline{\xi}-1)+\gamma p^{}_- \end{array}\right ]\  ,
\label{}
\end{align}
with 
\begin{align}
D=(s+iU\overline{\xi})(s+iU\overline{\xi}+\gamma)+U^2-iU\gamma\overline{\xi}\ .
\end{align}
It follows from  Eq. (\ref{av})  that \cite{com2}
\begin{align}
\overline{\tilde{ X}}(s)=[s+\gamma+2iU\overline{\xi}]/D
\label{fX}\ .
\end{align}
Inspection of the definition of the advanced  Green's function, Eq. (\ref{Gadd}), reveals that its telegraph-process average is independent of the time $t$, and is given by
\begin{align}
\overline{G^{a}_{dd}(\omega)}=i\overline{\tilde{X}(s=\Gamma+i[\omega-\epsilon-U\overline{\xi}])}\ .
\label{TNGa}
\end{align}
The telegraph-process averaging  over the energy fluxes involves averages of  products of  Green's functions with  $\epsilon^{}_{d}(t)$, Eq. (\ref{ed}), and its derivative. There are essentially two quantities to average, $\epsilon^{}_{d}(t)[G^{a}_{dd}(\omega,t)-G^{r}_{dd}(\omega,t)]$, and $-i\epsilon^{}_{d}(t)G^{<}_{dd}(t,t)$. 
Their calculation is quite similar to the one represented in detail above, and is relegated to Appendix \ref{average}.

\section{The stationary fluxes}
\label{stationary}

Once all histories of the telegraph-process events are averaged upon,  the physical expressions become independent of time.
It is illuminating to begin the analysis of this stationary limit  by inspecting the averages of the occupation on the dot, $Q_{d}$ [Eq. 
(\ref{Q})], and of the power absorbed from the telegraph-noise source, $P_{d}$ [Eq. (\ref{P})].
Using Eq. (\ref{TNGa}) in conjunction with Eq. (\ref{GLTT}) gives
\begin{align}
\overline{Q^{}_{d}}=
2\int \frac{d\omega}{2\pi}\frac{\Gamma^{}_{L}f^{}_{L}(\omega)+\Gamma^{}_{R}f^{}_{R}(\omega)}{\Gamma}{\rm Im}[\overline{G^{a}_{dd}(\omega)}]\ . 
\end{align}
This is  in fact the usual expression for the stationary occupation of a dot coupled to two  reservoirs that supply electrons. The first factor is the weighted average of the two Fermi distributions [which becomes simply $f_{L}(\omega)=f_{R}(\omega)\equiv f(\omega)$ for an un-biased dot].
The second factor is the density of states on the dot, with
\begin{align}
\overline{G^{a}_{dd}(\omega)} &=[
\overline{G^{r}_{dd}(\omega)}]^{\ast}_{}\nonumber\\
&=\frac{\omega-\epsilon^{}_{-}-i(\Gamma+\gamma)}{[\omega-\epsilon^{}_{+}-i\Gamma][\omega-\epsilon^{}_{-}-i(\Gamma+\gamma)]-\overline{{\cal E}^{2}}}\  .
\label{GAAV}
\end{align}
As seen, the telegraph noise turns our single resonance on the dot (centered around $\epsilon$ in the absence of it) into two resonances,  \cite{Gurvitz2016} centered essentially around $\epsilon_{+}$ and $\epsilon_{-}$, 
\begin{align}
\epsilon^{}_{\pm}=\epsilon\pm U \overline{\xi}\ .
\label{epm}
\end{align}
with disparate widths, essentially (i.e., small values of $U$) $\Gamma$ and $\Gamma+\gamma$.

The more interesting factor  is the ``effective electric field"  (squared), $\overline{{\cal E}^{2}}$, 
\begin{align}
\overline{{\cal E}^{2}}= U^{2}(1-\overline{\xi}^{2})=4U^{2}p^{}_{+}(T)p^{}_{-}(T)\ .
\label{E2}
\end{align}
This quantity, which vanishes when $T=0$ [see Eq. (\ref{overx})] determines the power absorbed by the junction.
Indeed, using Eq. (\ref{der}) in Eq. (\ref{P}) yields
\begin{align}
\overline{P^{}_{d}}=-\overline{{\cal E}^{2}}\frac{4\Gamma\gamma}{2\Gamma+\gamma}
\int \frac{d\omega}{2\pi}\frac{\Gamma^{}_{L}f^{}_{L}(\omega)+\Gamma^{}_{R}f^{}_{R}(\omega)}{\Gamma}{\rm Im}[{\cal F}^{a}_{}(\omega)]\ , 
\label{Pdav}
\end{align}
where the function ${\cal F}^{a}$, given in Eq. (\ref{F}) and reproduced here for clarity,  is
\begin{align}
{\cal F}^{a}_{}(\omega)&=[{\cal F}^{r}_{}(\omega)]^{\ast}
\nonumber\\
&=
\frac{1}{[\omega-\epsilon^{}_{+}-i\Gamma][\omega-\epsilon^{}_{-}-i(\Gamma+\gamma)]-\overline{{\cal E}^{2}}}\ .
\label{Frep}
\end{align}

\begin{figure}[htp]
\includegraphics[width=6cm]{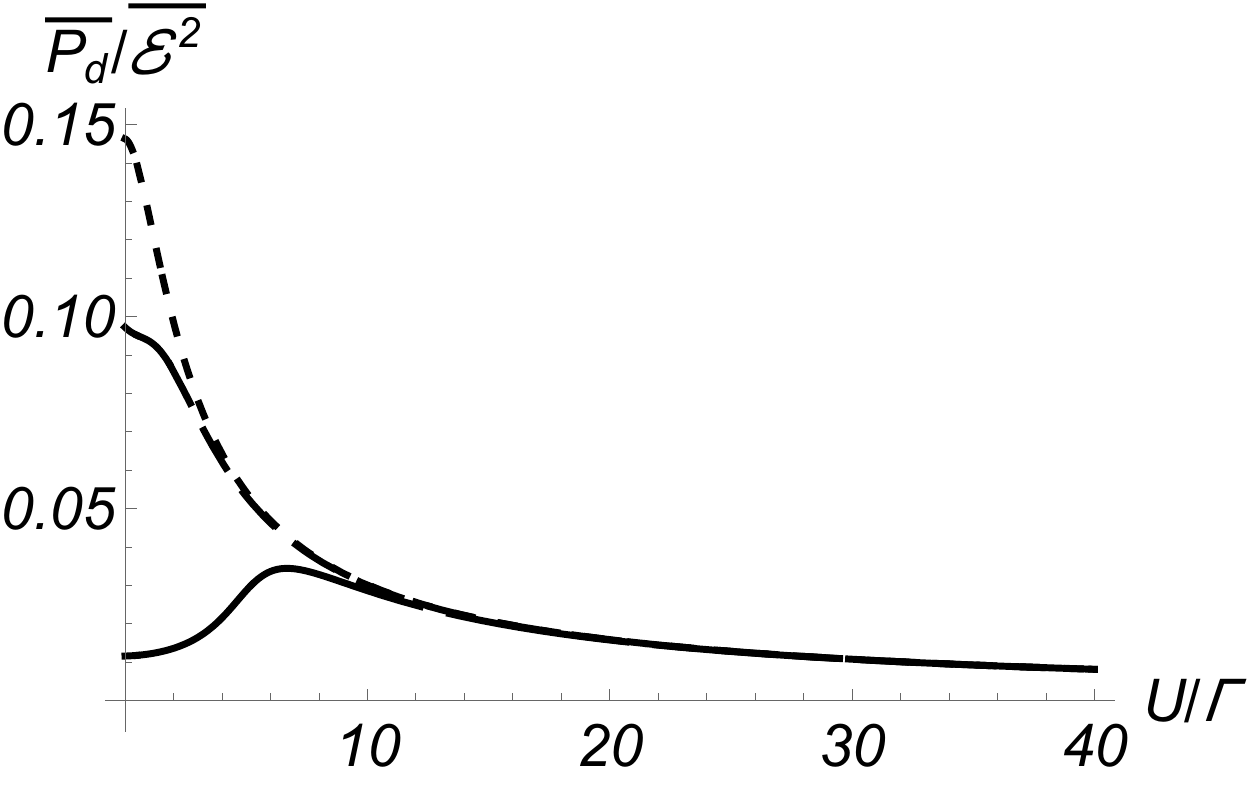}
\caption{The  power absorbed by the junction (in units of the effective electric field (squared) $\overline{{\cal E}^{2}}$, for identical electronic reservoirs (see text), $\overline{\xi}=0.5$ and  $\gamma=1$.  The different curves with increasing dashes are for $\epsilon-\mu= .1,\  1,\  5.\ $ (Energies are in units of $\Gamma$.)}
\label{Fig2}
\end{figure}

The expression for the power 
$\overline{P_{d}}$ possesses several remarkable properties. 
First, 
Eq. (\ref{Pdav}) yields a finite {\it positive}
power, as long as  the temperature of the noise source is finite, i.e., $|\overline{\xi}|\neq 1$.  This  implies that the junction absorbs energy from the source responsible for the telegraph noise.     For instance, 
when one confines oneself to the lowest order in $\overline{{\cal E}^{2}}$ (or, equivalently,  to order $U^{2}$),  then
\begin{align}
\lim_{\overline{{\cal E}^{2}}\rightarrow 0}{\rm Im}[{\cal F}^{a}_{}(\omega)]=\frac{1}{2\gamma}\frac{\partial}{\partial\omega}\ln\Big [\frac{(\omega-\epsilon)^{2}+\Gamma^{2}}{(\omega-\epsilon)^{2}+(\Gamma+\gamma)^{2}}\Big ]\ .
\end{align}
Inserting this expression into Eq. (\ref{Pdav}) and integrating by parts, one obtains
\begin{align}
\overline{P^{}_{d}}=\overline{{\cal E}^{2}}\frac{2}{2\Gamma+\gamma}
&\int \frac{d\omega}{2\pi}\Big (-\frac{\partial}{\partial\omega}[\Gamma^{}_{L}f^{}_{L}(\omega)+\Gamma^{}_{R}f^{}_{R}(\omega)]\Big )\nonumber\\
&\times
\ln\Big [\frac{(\omega-\epsilon)^{2}+(\Gamma+\gamma)^{2}}{(\omega-\epsilon)^{2}+\Gamma^{2}}\Big ]\ ,
\end{align}
which is obviously positive. In the more general case, but assuming  for simplicity \cite{com3} that 
the two electronic reservoirs are identical, i.e., $f_{L}(\omega)=f_{R}(\omega)\equiv f(\omega)$, one finds
\begin{align}
\label{integral}
&\int\frac{d\omega}{2\pi}f(\omega){\cal F}^{a}_{}(\omega)&\\
&=\frac{1}{\omega^{}_{+}-\omega^{}_{-}}\int\frac{d\omega}{2\pi}\Big (-\frac{\partial f(\omega)}{\partial\omega}\Big ) 
\ln\frac{\omega-\omega^{}_{+}}{\omega-\omega^{}_{-}}
\ ,
\nonumber
\end{align}
where
\begin{align}
\omega^{}_{\pm}&=\epsilon +i(\Gamma+\gamma/2)\pm \sqrt{U^{2}-(\gamma/2)^{2}-i\gamma U\overline{\xi}}\ .
\end{align}
Using this  expression,  Fig. \ref{Fig2} shows $\overline{P_{d}}/\overline{{\cal E}^{2}}$ as a function of 
$U/\Gamma$ for   $\overline{\xi}=0.5$ and three values of $\epsilon-\mu$, the bare energy on the dot relative to the common chemical potential in the leads, $\mu_{L}=\mu_{R}=\mu$ (the resonance width $\Gamma$ scales all energies).
At intermediate values of $U$ such that $U>\gamma$,  one observes a fast increase and a peak in $\overline{P_{d}}/\overline{{\cal E}^{2}}$  near $Up_{+}=\epsilon-\mu$. This peak arises as the chemical potential crosses the resonance at $\epsilon-Up_{+}$.
At large $U$ values,  the integral in Eq. (\ref{integral}) is well approximated by $-i\pi/U$, making the power {\it positive and   linear} in $U$, while it is {\it quadratic} in $U$ at small values. The curves in Fig. \ref{Fig2} are computed for zero electronic temperatures, i.e., for $[-\partial f(\omega)/\partial\omega]\rightarrow \delta (\omega)$. 
However, this approximation is also good at finite low temperatures, since the logarithmic function in Eq. (\ref{integral}) varies slowly with $\omega$. Indeed, numerical integration at higher electronic temperatures give similar qualitative results.
[The temperature of the noise source  is determined by $\overline{\xi}$, 
see Eq. (\ref{overx}).]

Second, the power absorbed by the junction necessitates  tunnel coupling with at least one lead. The reason for this is quite clear: an empty dot (which is the situation assumed for the decoupled junction) cannot absorb energy from a fluctuating electric field. Third, as noted, power is absorbed also when the junction is not biased, i.e., when $f_{L}(\omega)=f_{R}(\omega)\equiv f(\omega)$.
Finally, we note that the absorbed power vanishes when the mean rate $\gamma$ of  instantaneous jumps  vanishes.
This is in particular interesting in view of the fact that the telegraph noise does affect the density of states on the dot, Im[$\overline{G^{a}_{dd}(\omega)}]$ [see Eq. (\ref{GAAV})]
even when $\gamma=0$. \cite{comAA}

How is this flux of energy distributed between the two electronic reservoirs? To answer this question
we examine the various electronic currents. The average of the particle current over the telegraph processes 
[using Eq. (\ref{GAAV}) in Eq. (\ref{ILt})] takes the usual form of stationary particle current in a two-terminal junction,  \cite{Benenti2017}
\begin{align}
\overline{I^{}_{L}}=
\int\frac{d\omega}{2\pi}[f^{}_{R}(\omega)-f^{}_{L}(\omega)]{\cal T}(\omega)
\ , 
\label{ILW}
\end{align}
where
\begin{align}
{\cal T}(\omega)=4\Gamma^{}_{L}\Gamma^{}_{R}{\rm Im}[\overline{G^{a}_{dd}(\omega)}]/\Gamma
\label{T}
\end{align}
is the transmission through the junction; it
vanishes unless the junction is biased, and/or there is a temperature difference across it. The telegraph noise affects the density of states on the dot, and so modifies the transmission. \cite{Gurvitz2016}

The electronic energy fluxes, however, differ substantially from their ``canonical" forms in two-terminal junctions.
We find that the average of the energy flux associated with the left reservoir [see Eq. (\ref{ILEWt}) and the technical details given in Appendix \ref{average}] is
\begin{align}
\overline{I^{E}_{L}}=\int\frac{d\omega}{2\pi}[f^{}_{R}(\omega)-f^{}_{L}(\omega)]\omega{\cal T}(\omega)+\frac{\Gamma^{}_{L}}{\Gamma}\overline{P^{}_{d}}
\ .
\label{IELW}
\end{align}
The first term, which vanishes unless the junction is biased (by a voltage and/or temperature difference), is  indeed  the usual  energy current in a two-terminal electronic junction, with the transmission ${\cal T}(\omega)$ given in Eq. (\ref{T}). The second term comes from the power supplied by the source of the  telegraph processes. The sign of the first term depends on the bias, and/or  on  the temperature difference across the junction; this means that energy flux represented by the first term  can flow out or into the left reservoir. \cite{Benenti2017} As opposed, 
$\overline{P_{d}}$ is positive, which means that energy flows into the left reservoir from the source of the telegraph noise. Since 
$\overline{I^{E}_{R}}$ is given by Eq. (\ref{IELW}) with $L\Leftrightarrow R$, it follows that energy flows also into the right reservoir. Thus, the telegraph  noise supplies energies to both reservoirs, with the larger portion going into the more  strongly-coupled one.

Interestingly enough, the averages over the energy fluxes 
associated 
with the tunneling terms in the Hamiltonian,  $I^{E}_{{\rm tun},L(R)}$ [Eq. (\ref{ITLEt1})], vanish (as is also the case when the junction is subjected to an oscillatory field; this point is elaborated upon further  in Sec. \ref{disc}).
As 
a result,
one finds 
that
\begin{align}
\overline{I^{E}_{L}}+\overline{I^{E}_{R}}+\overline{I^{E}_{d}}=\overline{P^{}_{d}}\ ,
\end{align}
which is the stationary form of energy conservation in the junction [$I^{E}_{d}$ is the total average energy flux of the electrons on the dot, Eq. (\ref{IDE})].

\section{Discussion}
\label{disc}

Let us begin by summarizing our results and comparing them with the electronic fluxes derived in the presence of an ac electric field that acts on the dot. We considered a single-level quantum dot coupled to two leads, on which the electrons are subjected to a stochastic electric field
that imitates telegraph-noise  processes. Expressions for the electronic properties of this junction in the time domain can be found analytically within the Keldysh formalism for nonequilibrium Green's functions. This calculation is carried out without specifying the explicit time dependence of electric field; it thus pertains also for a junction in which a periodic ac electric field acts on electrons residing on the dot. \cite{Arrachea2016_94,Arrachea2017} We then focused on the specific time dependence that characterizes telegraph noise processes, and averaged the fluxes over the history of those processes, to obtain their stationary limits. In studies devoted to the effect of oscillatory fields, this step is replaced by an integration over a single  period of the ac field. It turns out that the energy currents associated with the tunneling terms in the Hamiltonian vanish when integrated over a period of the ac field; as found above, this is also the fate of these currents in the stationary limit of the telegraph noise.

Whereas  the telegraph noise affects only modestly the particle current by changing the density of states on the dot without inducing   any dramatic modifications, this is not the case with the energy currents. In contrast with the charge (or particle) current, that is established only when the junction is biased  (either by voltage or by temperature gradient), disparate energy currents do flow from the dot to the two leads even when  the two reservoirs are identical.  

We believe  that  investigations of the correlations of the  particle  and  energy currents will shed further light on this interesting problem.
This is because the average currents, either over  the telegraph processes or, in the case of an ac field over a period,  produce qualitatively similar results.
It may well be that the difference in the time dependence of the stochastic electric field and that of the oscillatory one will manifest itself in the correlations.

\begin{acknowledgments}
We thank L. Arrachea, S. Gurvitz, A. Nitzan,  R. Shekhter,  and J. Splettstoesser  for helpful discussions. OEW and AA are partially supported by  the Israel Science Foundation
(ISF), 
by the infrastructure program of Israel Ministry of Science and Technology under contract 3-11173, and by the Pazi Foundation. SD's research at the Bose Institute is supported by a Senior Scientist scheme of the Indian National Science Academy.

 \end{acknowledgments}

\begin{widetext}
 \appendix

\section{The Green's functions}
\label{GF}
In the time domain, the Dyson's equations  for the Green's functions        depend on two time arguments.
For instance, \cite{Jauho1994}
\begin{align}
G^{}_{{\bf k}d}(t,t')=\int dt^{}_{1}g^{}_{\bf k}(t,t^{}_{1})V^{}_{\bf k}G^{}_{dd}(t^{}_{1},t')
\ .
\end{align}
(The lowercase Green's functions pertain to the decoupled junction.)
The Green's function of the decoupled left reservoir is
 \begin{align}
g^{r(a)}_{\bf k}(t-t')&=\mp i\Theta (\pm t\mp t')\langle\{ c^{}_{\bf k}(t),c^{\dagger}_{\bf k}(t')\}\rangle
=\mp i\Theta (\pm t \mp t')e^{-i\epsilon^{}_{k}(t-t')}\ ,
\label{gr}
\end{align}
and
\begin{align}
g^{<}_{\bf k}(t-t')=if(\epsilon^{}_{k})e^{-i\epsilon^{}_{k}(t-t')}\ ,\ \ \ f(\epsilon^{}_{k})=\langle c^{\dagger}_{\bf k}c^{}_{\bf k}\rangle
\ .
\label{gl}
\end{align}
(The superscript $r(a)$  indicates
the  retarded (advanced) Green's  function and corresponds to the upper (lower) signs on the right hand-side.)
The Green's functions Eqs. (\ref{gr}) and (\ref{gl})  determine the self energies $\Sigma_{L}$ and $\Sigma^{E}_{L}$, Eqs. (\ref{SigL}) and (\ref{SigEL}), respectively.
When the densities of states in the reservoirs are assumed to be independent of the energy, i.e., in
the wide-band limit, 
 \cite{Jauho1994} then
\begin{align}
\sum_{k}|V^{}_{{\bf k}({\bf p})}|^{2}g^{r(a)}_{k(p)}(\omega)\approx \mp i\Gamma^{}_{L(R)}\ ,
\end{align}
where $
 \Gamma_{L(R)}=\pi|V^{}_{k_{\rm F}}|^{2}{\cal N}_{L(R)}$,
${\cal N}_{L(R)}$ is the constant  density of states of the left (right) lead. As a result, 
\begin{align}
&\Sigma^{r(a)}_{L(R)}(\omega)=\mp i\Gamma^{}_{L(R)}\ ,\ \ \ \Sigma^{<}_{L(R)}(\omega)=
2i\Gamma^{}_{L(R)}
f^{}_{L(R)}(\omega)
\ , \nonumber\\ \
&\Sigma^{Er(a)}_{L(R)}(\omega)=\mp i\omega\Gamma^{}_{L(R)}\ ,\ \ \  \Sigma^{E<}_{L(R)}(\omega)=2 i\omega\Gamma^{}_{L(R)}
 f^{}_{L(R)}(\omega)
\ ,
\label{WBL}
\end{align}
where
$f^{}_{L(R)}(\omega)$ [Eq. (\ref{Fermi})] 
is the Fermi distribution there.
 (Note that the self energies depend only on the time difference, and therefore are conveniently represented by their Fourier transforms.)

The Dyson equation for the Green's function on the dot reads
 \begin{align}
G^{}_{dd}(t,t')
&=g^{}_{d}(t,t')+\int dt^{}_{1}dt^{}_{2}g^{}_{d}(t,t^{}_{1})\Sigma^{}_{}(t^{}_{1},t^{}_{2})G^{}_{dd}(t^{}_{2},t')\ ,
\label{DG}
\end{align}
where
\begin{align}
\Sigma(t,t')=\Sigma^{}_{L}(t,t')+\Sigma^{}_{R}(t,t')\ , \ \ \Gamma=\Gamma^{}_{L}+\Gamma^{}_{R}\ .
\end{align}
The Green's function of the decoupled dot is
\begin{align}
g^{r(a)}_{d}(t,t')=\mp i\Theta (\pm t\mp t')e^{ -i\int_{t'}^{t}dt^{}_{1}\epsilon^{}_{d}(t^{}_{1})}\ ,
\label{grad}
\end{align}
and $g^{<}_{d}=0$, since it is assumed that the dot is empty in the decoupled junction.
Solving Eq. (\ref{DG})  for the retarded (advanced) Green's function gives \cite{Odashima2017}
\begin{align}
G^{r(a)}_{dd}(t,t')&=\mp i \Theta (\pm t\mp t')
e^{-i\int_{t'}^{t}dt^{}_{1}\epsilon^{}_{d}(t^{}_{1})\mp \Gamma (t-t')}
\ .
\label{GDRA}
\end{align}
The Fourier transforms of these functions, 
\begin{align}
G^{r}_{dd}(\omega ,t)=\int dt'e^{i\omega(t-t')}G^{r}_{dd}(t,t')\ ,\ \ 
G^{a}_{dd}(\omega ,t)=\int dt'e^{-i\omega(t-t')}G^{a}_{dd}(t',t)
\ ,
\end{align}
lead to Eq. (\ref{GDRAW}) in the main text.
Inserting these solutions into the Dyson equation (\ref{DG}) for the lesser Green's function on the dot yields
\begin{align}
G^{<}_{dd}(t,t)&=\int\frac{d\omega}{2\pi}\Sigma^{<}_{}(\omega)\int^{t}dt^{}_{1}\int^{t}dt^{}_{2}e^{\Gamma(t^{}_{1}+t^{}_{2}-2t)}\
 e^{i\omega(t^{}_{2}-t^{}_{1})}e^{i\int_{t^{}_{2}}^{t_{1}^{}}d\tau\epsilon^{}_{d}(\tau)}\ .
\label{GLTTA}
\end{align}
By changing the double integration,
\begin{align}
&\int^{t} dt^{}_{1}\int^{t}dt^{}_{2}F(t^{}_{1},t^{}_{2})=
 \int^{t} dt^{}_{1}\int^{t^{}_{1}}dt^{}_{2}[F(t^{}_{1},t^{}_{2} )+F(t^{}_{2},t^{}_{1})]\ ,
 \end{align}
 and using Eq.
(\ref{GDRAW}) one derives Eq. (\ref{GLTT}).

The rate of energy associated with the left reservoir [see Eqs. (\ref{ILE}) and (\ref{ILEd})]  involves the expression
\begin{align}
&\int dt'[\Sigma^{Er}_{L}(t,t')G^{<}_{dd}(t',t)-G^{<}_{dd}(t,t')\Sigma^{Ea}_{L}(t',t)]=-\Gamma^{}_{L}\int d\tau\delta (\tau)
\frac{\partial}{\partial \tau}[G^{<}_{dd}(t,t-\tau)-G^{<}_{dd}(t-\tau,t)]\ ,
\end{align}
where we have used Eqs. (\ref{WBL}).
Inserting here
\begin{align}
&G^{<}_{dd}(t,t')=\int\frac{d\omega}{2\pi}e^{-i\omega(t-t')}G^{r}_{dd}(\omega,t)\Sigma^{<}_{}(\omega)G^{a}(\omega,t')\ ,
\label{GLESW}
\end{align}
[$\Sigma^{<}_{}(\omega)$ is given in Eq. (\ref{siglew})] and carrying out the derivatives
exploiting Eqs. (\ref{GDRA})  and (\ref{GDRAW}),  results in the last term on the right hand-side of Eq. (\ref{ILEWt}).

The last term in Eq. (\ref{ITLE}) for the energy rate associated with the tunneling between the dot and the left reservoir is calculated from the Dyson equation
\begin{align}
G^{}_{{\bf k}{\bf p}}(t,t')&=V^{}_{\bf k}V^{\ast}_{\bf p}\int dt^{}_{1}\int dt^{}_{2}g^{}_{k}(t,t^{}_{1})G^{}_{dd}(t^{}_{1},t^{}_{2})g^{}_{p}(t^{}_{2},t')\ .
\label{dk}
\end{align}
With the definition of the self energy Eq. (\ref{SigL}) (and the analogous one for $\Sigma_{R}$), this term in Eq. (\ref{ITLE}) becomes
\begin{align}
&\int dt'\int dt''[\Sigma^{}_{L}(t,t')G^{}_{dd}(t',t'')
\Sigma^{}_{R}(t'',t)
-\Sigma^{}_{R}(t,t')G^{}_{dd}(t',t'')\Sigma^{}_{L}(t'',t)]^{<}_{}\ .
\label{kp}
\end{align}
It remains to apply the Langreth rules to the product of Green's functions; the result is the last term in Eq. (\ref{ITLEt1}).

\section{Details of the average over the telegraph process}
\label{average}

As mentioned in Sec. \ref{RTN},  the averages of the energy fluxes are determined  by  averaging over products of  Green's functions with  $\epsilon^{}_{d}(t)$, Eq. (\ref{ed}), and its derivative. Using Eq.  (\ref{x}) and  the definition
\begin{align}
Y(t,t',t'')=\xi(t)X(t',t'')\ ,\ \ {\rm for}\ \ t>t'>t''\ ,
\label{y}
\end{align}
one can write
\begin{align}
\epsilon^{}_{d}(t)G^{a}_{dd}(\omega,t)&=
i\int^{t}dt'(e^{-i(\omega-\epsilon^{}_{+}-i\Gamma)(t-t')}[\epsilon X(t,t')+UY(t,t,t')]\ .
\label{eG}
\end{align}
The inverse Laplace transform of  Eq. (\ref{fX}) yields
\begin{align}
\overline{ X(t,t')}&=\overline{ X(t-t')}=\frac{1}{2}[e^{s^{}_{+}(t-t')}+e^{s^{}_{-}(t-t')}]+\frac{\gamma/2+iU\overline{\xi}}{2u}[e^{s^{}_{+}(t-t')}-e^{s^{}_{-}(t-t')}]\ ,
\label{Xt}
\end{align}
with
\begin{align}
u=\sqrt{(\frac{\gamma}{2}+iU\overline{\xi})^{2}-\overline{{\cal E}^{2}}}\ ,\ \ \ 
{\rm and}\ \ \ 
s^{}_{\pm}=-\frac{\gamma}{2}\pm u\ .
\end{align}
[$\overline{{\cal E}^{2}}$ is defined in Eq. (\ref{E2}).] 
Similarly, 
\begin{align}
&\overline{ Y(t,t',t'')}
\equiv \overline{Y(t-t',t-t'')}=\overline{\xi} \overline{X(t'-t'')}
+\frac{iU(1-\overline{\xi}^{2})}{2u}[e^{s^{}_{+}(t'-t'')}-e^{s^{}_{-}(t'-t'')}]e^{-\gamma(t-t')}\ . 
\label{Yt}
\end{align}
Inserting Eqs. (\ref{Xt}) and (\ref{Yt}) into the average of Eq. (\ref{eG}) yields 
\begin{align}
\overline{\epsilon^{}_{d}(t)G^{a}_{dd}(\omega ,t)}&=\overline{\epsilon^{}_{d}(t)[G^{r}_{dd}(\omega ,t)]^{\ast}}
=\epsilon^{}_{+}\overline{G^{a}_{dd}(\omega)}+
\overline{{\cal E}^{2}}{\cal F}^{a}_{}(\omega)\ ,
\label{eGav}
\end{align}
where the  function ${\cal F}^{a}(\omega)$ is 
\begin{align}
{\cal F}^{a}_{}(\omega)&=[{\cal F}^{r}_{}(\omega)]^{\ast}
=
\Big [[\omega-\epsilon^{}_{+}-i\Gamma][\omega-\epsilon^{}_{-}-i(\Gamma+\gamma)]-\overline{{\cal E}^{2}}\Big ]^{-1}_{}\ , 
\label{F}
\end{align}
with $\epsilon_{\pm}$ given in Eq. (\ref{epm}).

The second quantity to average is
$-i\epsilon^{}_{d}(t)G^{<}_{dd}(t,t)$.  Inserting the definitions (\ref{x}) and (\ref{y}) into Eq. (\ref{GLTT}) gives
\begin{align}
&i\epsilon^{}_{d}(t)G^{<}_{dd}(t,t)=\int\frac{d\omega}{2\pi}\Sigma^{<}_{}(\omega)
\int^{t}dt'e^{-2\Gamma(t-t')}
\int^{t'}
dt''\Big (ie^{-i(\omega-\epsilon^{}_{+}-i\Gamma)(t'-t'')}[\epsilon X(t'-t'')+UY(t,t',t'')]-{\rm cc}\Big )\ .
\end{align}
The average over the telegraph process of this expression is obtained upon using Eq. (\ref{Xt}) and (\ref{Yt}), 
\begin{align}
i\overline{\epsilon^{}_{d}(t)G^{<}_{dd}(t,t)}
&=\int\frac{d\omega}{2\pi}\Sigma^{<}_{}(\omega)\Big (\frac{\omega-i\Gamma}{2\Gamma}\overline{G^{a}_{dd}(\omega)}
-\overline{{\cal E}^{2}}\frac{\gamma}{2\Gamma(2\Gamma+\gamma)}{\cal F}^{a}_{}(\omega)-{\rm cc}\Big )\ .
\label{3t}
\end{align}
Note the identity
\begin{align}
(\omega-i\Gamma)\overline{G^{a}_{dd}(\omega)}-{\rm cc}
=\epsilon^{}_{+}\overline{G^{a}_{dd}(\omega)}+\overline{{\cal E}^{2}}{\cal F}^{a}_{}(\omega)-{\rm cc}\ ,
\end{align}
which turns the average Eq. (\ref{3t}) into
\begin{align}
i\overline{\epsilon^{}_{d}(t)G^{<}_{dd}(t,t)}
&=\int\frac{d\omega}{2\pi}\Sigma^{<}_{}(\omega)
\Big [\frac{\epsilon^{}_{+}}{2\Gamma}\overline{G^{a}_{dd}(\omega)}
+\frac{\overline{{\cal E}^{2}}}{2\Gamma+\gamma}{\cal F}^{a}_{}(\omega)-{\rm cc}\Big ]\ .
\end{align}
Finally, the average of the product of $d\epsilon_{d}(t)/dt$  with $G^{<}$ is
\begin{align}
i\overline{\frac{d\epsilon^{}_{d}(t)}{dt}G^{<}_{dd}(t,t)}=&\int\frac{d\omega}{2\pi}\Sigma^{<}_{}(\omega)
\int^{t}dt'e^{-2\Gamma(t-t')}
\int^{t'}
dt''\Big (ie^{-i(\omega-\epsilon-i\Gamma)(t'-t'')}U\overline{\frac{d Y(t,t',t'')}{dt}}-{\rm cc}\Big )\nonumber\\
&=-\int\frac{d\omega}{2
\pi}\Sigma^{<}_{}(\omega)\Big [\frac{\gamma\overline{{\cal E}^{2}}}{
2\Gamma+\gamma}{\cal F}^{a}_{}(\omega)-{\rm cc}\Big ]\ .
\label{der}
\end{align}

\end{widetext}


\begin{thebibliography}{99}
\bibitem{com3}
This is not a drastic approximation: in the general case, the derivative $\partial f(\omega)/\partial\omega$ is replaced by $\partial [\Gamma{}_{L}f^{}_{L}(\omega)+\Gamma^{}_{R}f^{}_{R}(\omega)]/(\Gamma\partial\omega)$.
\bibitem{comAA}
In this limit $\omega_{\pm}=\epsilon \pm U+i\Gamma$,   corresponding to the two possible initial dot energies $\epsilon\pm U$. At  $\gamma=0$ the dot energy remains at its initial value, but the telegraph noise averaging yields two resonances.
\end{thebibliography}
\end{document}